\documentclass[amsmath,amssymb,pra,final,showpacs,twocolumn]{revtex4-1}
 
\usepackage{bm,hyphenat,xspace}
\usepackage{graphicx,epsfig}

\newcommand {\mbf}[1]{{\mathbf{#1}}}

\newcommand {\mcj}{\mathcal{J}}
\newcommand {\mcu}{\mathcal{U}}

\newcommand{\He}{${}^4\mathrm{He}$ }

\begin{document}

\title {Four-body  system of ${}^4\mathrm{He}$ atoms: Dimer-dimer scattering}
  
\author{A.~Deltuva} 
\affiliation
{Institute of Theoretical Physics and Astronomy, 
Vilnius University, Saul\.etekio al. 3, LT-10257 Vilnius, Lithuania
}

\received{February x, 2022} 

\begin{abstract}
  The strong short-range repulsion, characteristic to realistic interatomic potentials, complicates
  the description of  weakly-bound few-body systems such as those of \He atoms.
  The present work proposes an approach for solving this problem and applies it to a realistic
  system of four ${}^4\mathrm{He}$ atoms. The potential is gradually softened such that
  rigorous  four-body equations for bound and scattering states can be accurately solved in the
  momentum-space framework, and the results are extrapolated back to the limit of the original potential.
  Linear correlations between three- and four body quantities are observed, and the accuracy 
  of the procedure is improved by extrapolating in one of the three-body quantities.
  Results for the ${}^4\mathrm{He}$  tetramer ground and excited state binding energies and
  atom-trimer scattering agree well with at least some of earlier determinations and shed light on
  the existing disagreements. An additional case of the Phillips correlation line is established
  for the dimer-dimer scattering length. The trimer production rate via the ultracold two-dimer collisions
  is estimated, it exhibits significant finite-range effects despite the weak binding of the dimer.
\end{abstract}

 \maketitle

\section{Introduction \label{sec:intro}}

Cold \He atoms constitute one of the simplest quantum systems exhibiting the phenomenon of
the Efimov physics \cite{efimov:plb,kunitski:15a}; see Refs.~\cite{naidon:rev,kievsky:21arnps}
for  recent reviews.
In contrast to atomic systems of alkali metals, no external fine-tuning is needed since the 
interatomic \He  interaction supports a single shallow dimer bound state in the $S$-wave.
Consequently, the two-atom scattering length is large compared to the interaction or
effective range, ensuring conditions for the realization of the Efimov physics.
It manifests itself by the existence few-body states. In systems consisting of $N=3$ (4) atoms 
of \He this leads  to two bound trimer (tetramer) states, a more tightly bound ground state and
a shallow excited state located near the two-cluster breakup threshold.

The system of three \He atoms has been  studied in a large number of works employing 
 realistic interaction models, and a good agreement between different theoretical methods 
has been achieved, not only for binding energies but also for the atom-dimer scattering length
\cite{blume:00a,barletta:01a,lazauskas:he,kolganova:09a,roudnev:11a,das:11a,hiyama:12a,deltuva:15g}. 
Several existing calculations 
for the ground state \He tetramer also agree well, but significant differences show up in the 
excited tetramer binding energy \cite{blume:00a,lazauskas:he,das:11a,hiyama:12a}.
The latter is correlated with the atom-trimer scattering length, the few existing results
being in sizable disagreement, thereby calling for further studies with alternative methods.
Furthermore, all available four-atom calculations with realistic \He potentials are limited
to energies very close to the atom-trimer threshold, with no predictions at higher energy where
inelastic scattering channels become open, enabling rearrangement reactions such as the dimer-trimer 
conversion.

The most important reason for the above-mentioned disagreements and limitations is the form of the 
realistic interaction between two \He atoms, namely, the weakly attractive van der Waals tail 
and the very strong repulsion at short distance. For spatially extended weakly-bound or scattering states
the physical observables result from a very subtle interplay of those two features, rendering the
numerical solution very sensitive to fine details and eventually leading to a significant accuracy loss.
For example, to overcome these difficulties when solving the coordinate space Faddeev-Yakubovsky equations
Ref.~\cite{lazauskas:he}
 had to impose additional boundary conditions in the hard-core region and apply extrapolation in the grid size.
The momentum-space method based on the Alt, Grassberger, and Sandhas (AGS) equations \cite{grassberger:67}
for the four-particle transition operators, although very efficient in realistic four-nucleon reaction
calculations \cite{deltuva:14a}, has not yet been successfully applied to the problem of four \He atoms
with realistic potential models.

The aim of the present work is to develop the momentum-space method for the four-body calculation 
in the multichannel regime with a realistic interatomic \He potential, to show its reliability and evaluate 
the complex dimer-dimer scattering length. The idea is to reduce gradually the strength of the short-range
repulsion, such that accurate solutions of the integral equations for transition operators or wave-function
components can be obtained, and then perform extrapolation of the results back to the original potential.

Section II introduces the scheme for reducing the short-range repulsion, whereas Sec. III demonstrates its
validity in the three-body system. Section IV shortly recalls the equations for the four-body system
together with the essential aspects of calculations. Section V presents results for tetramer
binding energies and atom-trimer and dimer-dimer scattering.
Summary and conclusions are collected in Sec. VI.

 \section{Transformation of the potential \label{sec:2b}}

All realistic interatomic \He potentials have in common the weakly attractive van der Waals tail 
and  strong repulsion at short distances $r < 2.5$ \AA. The most widely used parametrization
is the LM2M2 by Aziz and Slaman \cite{aziz}, it will be adopted also in the present work
with the $\hbar^2/m=12.11928$ K\AA$^2$ value recommended in Ref.~\cite{roudnev:11a}, slightly
different from the approximation  $\hbar^2/m=12.12$ K\AA$^2$ used Ref.~\cite{lazauskas:he}
and many other works, where $m$ is the mass of the \He atom.
Anyway, as Refs.~\cite{roudnev:11a,hiyama:12a} demonstrated, the changes in three- and four-body
results due to different $\hbar^2/m$ are small and can be accounted for by  a simple
perturbative correction.

In nuclear physics, the short-range repulsion in the two-nucleon potentials can be softened by 
the similarity renormalization group (SRG) method \cite{bogner:07b}, which is a unitary
transformation decoupling low- and high-momentum components while
preserving the deuteron binding and two-nucleon phase shifts. However,
in the interatomic \He force the relative impact of the short-range repulsion is considerably
stronger than in the nuclear force, precluding the  application of the SRG or
similar method; no successful attempt is reported so far.

\begin{figure}[!]
\begin{center}
\includegraphics[scale=0.64]{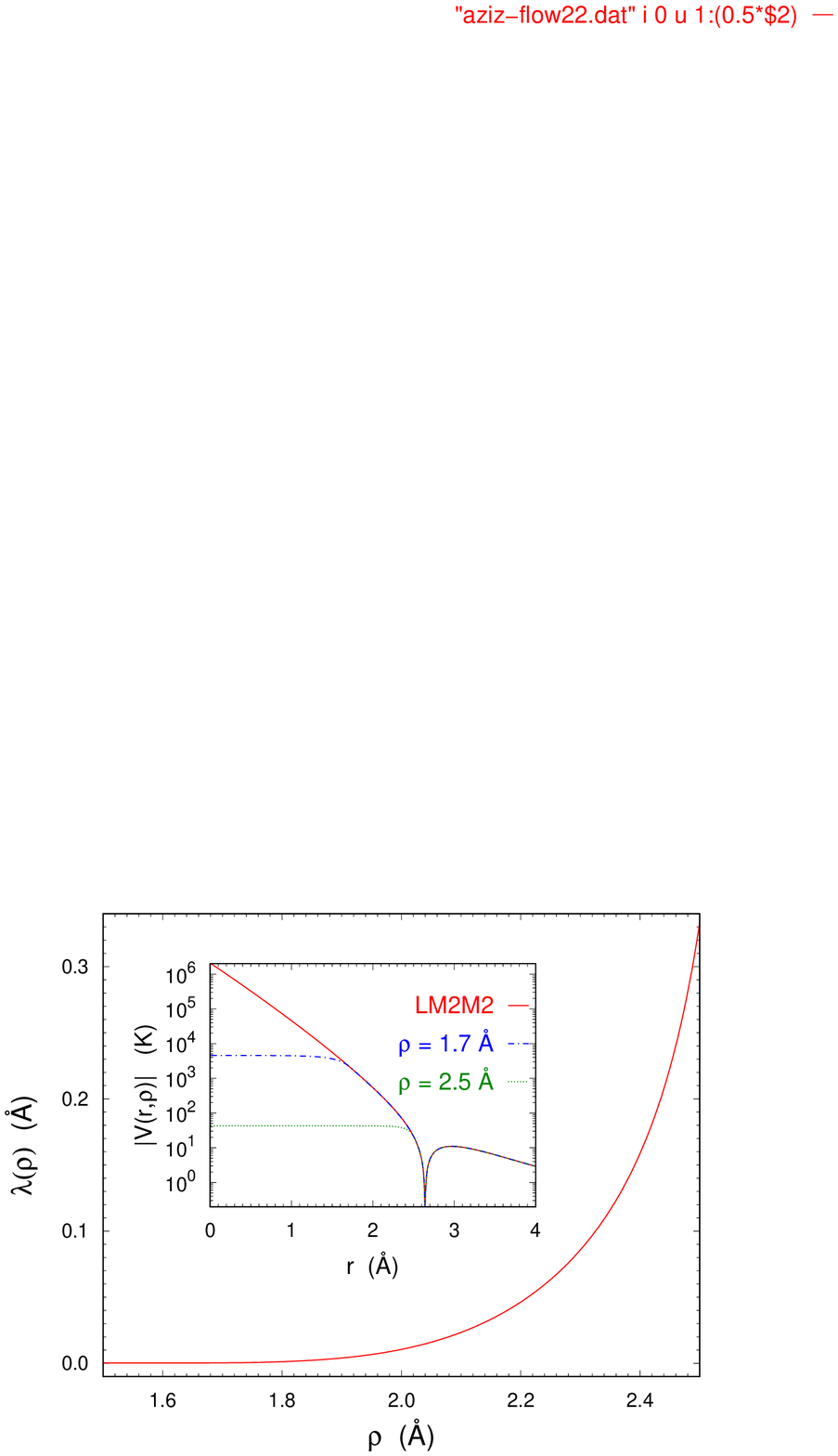}
\end{center}
\caption{\label{fig:lam} (Color online)
  The function $\lambda(\rho)$  controlling the momentum-dependent reduction in the potential
  (\ref{eq:vpr}). The inset compares the original LM2M2 potential (solid curve) and the
  auxiliary potential (\ref{eq:vrr}) at $\rho = 1.7$ and 2.5 \AA, given by
  dashed-dotted and dotted curves, respectively.
}
\end{figure}

The present work will therefore use a different strategy. Following Ref.~\cite{deltuva:15g},
an auxiliary potential $V(r,\rho)$ with
the strength of the short-range repulsion  reduced at distances $r < \rho$  is defined as
\begin{gather} \label{eq:vrr}
V(r,\rho) =  \Theta(\rho-r) V(\rho)[2 - e^{(\rho-r)\kappa(\rho)} ] 
  +   \Theta(r-\rho)V(r).
\end{gather}
Here $V(r)$ is the original potential, $\Theta(x)$ is the step function equal to 1 (0) for positive (negative)
argument $x$,
while $\kappa(\rho) = V'(r)/V(r)|_{r=\rho}$
ensures the smoothness of the auxiliary potential by 
the continuity of itself and its first derivative. The comparison of the original LM2M2 potential
with  $V(r,\rho)$ is shown in the inset of Fig.~\ref{fig:lam}. While the LM2M2 potential rapidly increases
towards $r=0$,  $V(r,\rho)$ remains quite flat for $r<\rho$.

For momentum-space partial-wave calculations the potential has to be transformed into the corresponding
representation, i.e.,
\begin{gather} \label{eq:Vpr}
\langle p' | V_L(\rho) | p  \rangle = \frac{2}{\pi} \int_0^{\infty} j_L(p'r)  \,V(r,\rho)  \,
  j_L(pr) \, r^2 dr,
\end{gather}
where $p$ ($p'$) is the initial (final) relative two-particle momentum, $L$ the orbital angular momentum,
and $j_L(x)$ the spherical Bessel function. The upper integration limit is formally infinite,
but Ref.~\cite{deltuva:15g} showed that at least 5 digit accuracy for observables is achieved
with a finite upper limit of 75 \AA; the present work uses 100 \AA{} which is fully sufficient,
since the integral equation formulation of the scattering theory in momentum space 
includes the asymptotic boundary conditions implicitly.
Reference~\cite{deltuva:15g} investigated also the $\rho$-dependence of the three-body binding energy and 
atom-dimer phase shifts and demonstrated that within the 5 digit accuracy the results remain unchanged
for $\rho$ ranging from 0 to 1.7 \AA. This means that practically there is no penetration into the
$\rho \le 1.7$ \AA{}  region even if the potential is reduced by more than two orders of magnitude at $r=0$,
as the  inset of Fig.~\ref{fig:lam} indicates. On the other hand, it is quite obvious that the proposed
approach will fail beyond $\rho \approx 2.5$ \AA, where the barrier height and the minimum
depth become of comparable size.
However, already beyond $\rho = 1.7$ \AA{} a further increase of $\rho$ reduces the short-range
repulsion to the extent that
it becomes insufficient to preserve the fine-tuned balance with the longer-range attraction.
As a consequence, the binding energies of few-body bound states start to increase with
$\rho$ for $\rho > 1.7$ \AA. 
To keep the same energy scale for all considered $\rho$, I choose to fix the dimer binding energy
$B_2= 1.3094$ mK. For this, a further modification of the potential is needed that removes  the
overbinding. In effective field theories this is achieved by a repulsive short-range counterterm,
but such a solution is unwanted in the present context as it would increase again the short-range repulsion.
Therefore I propose a different approach, rendering the potential (\ref{eq:Vpr}) weaker.
There are many ways to achieve this goal, e.g., to rescale the potential (\ref{eq:Vpr})
by a constant or momentum-dependent factor, or even by a combination of them.
My primary choice is the modified potential
\begin{gather} \label{eq:vpr}
  \langle p' | v_L(\rho) | p  \rangle  =
  e^{-[\lambda(\rho)p']^2} \, \langle p' | V_L(\rho) | p  \rangle \, e^{-[\lambda(\rho)p]^2}
\end{gather}
where the function $\lambda(\rho)$ controls the momentum-dependent reduction. It
is determined by fitting the  dimer binding energy $B_2= 1.3094$ mK in the $L=0$
partial wave, but the same  $\lambda(\rho)$ is used for all partial waves. Thus, none of phase shifts
are explicitly fitted and therefore may deviate from their original values. Nevertheless, this deviation
is not really important as long as no new bound or resonant states appear.
Since $\lambda(\rho)$ smoothly approaches zero with decreasing $\rho$ and vanishes for $\rho < 1.7$ \AA{}
as shown in Fig.~\ref{fig:lam},
the potential (\ref{eq:vpr}) smoothly converges towards the original potential and all observables should
do so as well. 
Further important advantage of the potential (\ref{eq:vpr}) is the suppression of high-momentum components
that is favorable for the stability of numerical calculations. Low-momentum components corresponding
to the long-range tail are barely affected. Though  the
potential (\ref{eq:vpr}) becomes   nonlocal in the coordinate space,
this presents no additional difficulty for momentum-space calculations.

In summary, few-body equations have to be solved for a series of potentials (\ref{eq:vpr}) with different values
of the softening
parameter $\rho$, large enough for numerically stable solution, and the obtained results extrapolated towards
smaller $\rho$ values until independence of $\rho$ is achieved. For brevity, the method will be referred to
as ``softening and extrapolation'' (SE).

\section{Test case: Three-body system \label{sec:3b}}

The bound state and scattering
problem of three \He atoms with realistic interaction in the momentum-space framework was solved 
using the original potential or the one of Eq.~(\ref{eq:Vpr}) with $\rho \le 1.7$ \AA{} 
\cite{deltuva:15g}. Thus,  the SE  is not needed to obtain physical results, 
but it is a good test case to demonstrate its validity.

The considered three-body observables are the atom-dimer scattering length $A_{12}$, and 
binding energies for the trimer ground and excited states, $B_3$ and $B_3^*$, respectively.
More details of calculations can be found in  Ref.~\cite{deltuva:15g},
the only difference here being the updated $\hbar^2/m$ value and modified potential (\ref{eq:vpr}).
At $\rho \le 1.7$ \AA{} the predictions
$A_{12} = 115.39$ \AA,  $B_3 = 126.50$ mK and $B_3^* = 2.2784$ mK
agree perfectly with the benchmark calculation  \cite{roudnev:11a}.

\begin{figure}[!]
\begin{center}
\includegraphics[scale=0.64]{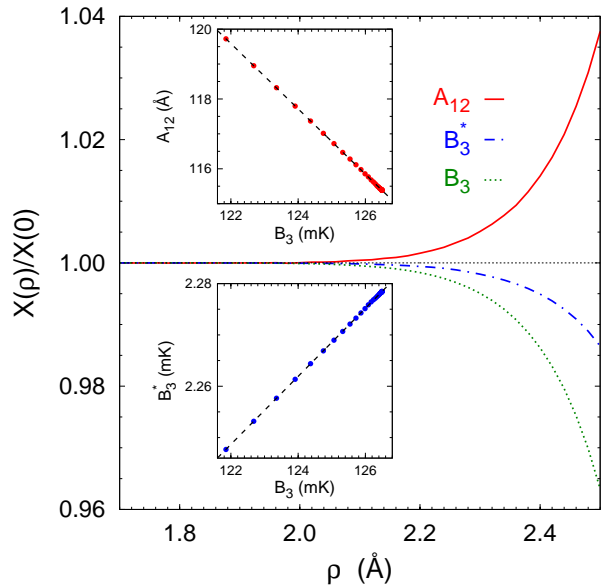}
\end{center}
\caption{\label{fig:b3} (Color online)
 Atom-dimer scattering length and
 trimer ground and excited state binding energies,
all normalized by the respective predictions obtained with the original LM2M2 potential,
 are shown  as  functions of the softening parameter.
Insets display correlations between trimer binding energies and atom-dimer scattering length.
}
\end{figure}

Figure \ref{fig:b3} presents the dependence of the atom-dimer scattering length and
trimer binding energies on the softening parameter $\rho$.
The quantities are normalized by their original values taken at $\rho = 0$  and 
listed above, in order to keep a single scale.
The ratios start deviating from the unity above $\rho = 2.0$ \AA, and reach nearly  4\% difference
at $\rho = 2.5$ \AA.
As functions of $\rho$, the deviations grow with  increasing rate, which makes the extrapolation
in $\rho$ problematic if data points near the plateau regime are not available. 
However, a closer inspection reveals that for all three quantities in Fig.~\ref{fig:b3} the shape
of deviation is very similar.
The consequence of this feature are linear correlations between trimer binding energies and
atom-dimer scattering length; two pairs, $A_{12}(B_3)$ and $B_3^*(B_3)$,
are shown in the insets of Fig.~\ref{fig:b3}, and
 the third pair $A_{12}(B_3^*)$ correlates equally well.
These correlations are expected in the context of Efimov physics, they are known as Tjon and Phillips
lines also in nuclear physics \cite{naidon:rev,kievsky:21arnps}.

The observed correlations suggest an alternative way of extrapolation by changing
the extrapolation variable, instead of $\rho$ considering 
one of three-body quantities, for example $B_3(\rho)$, given that it can be reliably calculated for small $\rho$,
reproducing the original limit $B_3 = B_3(0)$. 
Other three-body quantities like $A_{12}(\rho)$ and $B_3^*(\rho)$ show (nearly) linear dependence on $B_3(\rho)$,
and therefore, even if calculated only at $\rho$ values beyond the plateau region,
as functions of $B_3(\rho)$ can  be reliably extrapolated to the point $B_3(0)$,
thereby yielding $A_{12}(0)$  and $B_3^*(0)$ estimations.
The lines in the insets of Fig.~\ref{fig:b3} are linear fits to data in the regime
2.3 \AA{} $ \le \rho \le 2.5$ \AA{}. The extrapolation towards the $B_3(0)$ value reproduces
the $A_{12}(0)$  and $B_3^*(0)$ within 0.01\% accuracy, confirming the reliability of the 
proposed extrapolation approach.

At a first glance it may appear that shapes of observable deviation in Fig.~\ref{fig:b3}
and of $\lambda(\rho)$ in Fig.~\ref{fig:lam} are similar, suggesting  $\lambda(\rho)$ as a suitable
extrapolation variable. However, this is not true since in the regime 1.7 \AA{} $ < \rho < 2.0$ \AA{}
there are small but visible changes in $\lambda(\rho)$, in contrast to
$A_{12}(\rho)$, $B_3(\rho)$,  and $B_3^*(\rho)$. Consequently, their dependence on $\lambda(\rho)$ is nearly linear
at $\rho > 2.3$ \AA, but bends for smaller $\rho$, preventing reliable extrapolation.
One could find perhaps a better behaving function of $\rho$ like $(\rho-\rho_0)^n$  or 
$[\lambda(\rho)]^n$ that could be more suitable as extrapolation
variable, but it is not trivial to achieve the linearity as good as  with $B_3(\rho)$.

\section{Four-body equations \label{sec:eq}}

The momentum-space integral-equation approach to the four-body problem starts
with the two-body transition operator
\begin{equation} \label{eq:t}
t= v + v G_0 t
\end{equation}
that sums up the respective pair interaction $v$ to all orders. The dependence on the
available energy $E$ arises via the free resolvent
$G_0 = (E+i0-H_0)^{-1}$  with kinetic energy operator $H_0$.
The bound state energy in the system of four identical bosons 
can be obtained from symmetrized Faddeev-Yakubovsky equations \cite{yakubovsky:67}
for wave-function components
\begin{subequations} \label{eq:fy}
\begin{align}  \label{eq:fy1}
|\psi_1 \rangle = {}&  G_0 t P_1 [(1+P_{34})|\psi_1 \rangle + |\psi_2 \rangle], 
\\ \label{eq:fy2}
|\psi_2 \rangle = {}&  G_0 t P_2 [(1+P_{34})|\psi_1 \rangle + |\psi_2 \rangle], 
\end{align}
\end{subequations}
where $t$ acts within pair (12) and $P_{ab}$ interchanges particles $a$ and $b$,
while $P_1 =  P_{12}\, P_{23} + P_{13}\, P_{23}$ and $P_2 =  P_{13}\, P_{24} $.

The scattering processes are described using AGS equations \cite{grassberger:67}
for four-particle transition operators $\mcu_{\beta\alpha}$. In the symmetrized form
they read
\begin{subequations} \label{eq:U}
\begin{align}  
\mcu_{11}  = {}&  P_{34} (G_0  t  G_0)^{-1}  
+ P_{34}  U_1 G_0  t G_0  \mcu_{11}   + U_2 G_0  t G_0  \mcu_{21} , 
\label{eq:U11} \\  
\mcu_{21}  = {}&  (1 + P_{34}) (G_0  t  G_0)^{-1}  
+ (1 + P_{34}) U_1 G_0  t  G_0  \mcu_{11} , \label{eq:U21} \\
\mcu_{12}  = {}&  (G_0  t  G_0)^{-1}  
+ P_{34}  U_1 G_0  t G_0  \mcu_{12}  + U_2 G_0  t G_0  \mcu_{22} , 
\label{eq:U12} \\  
\mcu_{22}  = {}& (1 + P_{34}) U_1 G_0  t  G_0  \mcu_{12} . \label{eq:U22}
\end{align}
\end{subequations}
Here the subscripts $\alpha,\beta= 1$ (2) label the 3+1 (2+2) clustering,
while 
\begin{equation} \label{eq:U3}
U_{\alpha} =  P_\alpha G_0^{-1} + P_\alpha  t G_0  U_{\alpha}
\end{equation}
are the  3+1 or 2+2 subsystem transition operators.

Physical transition amplitudes for two-cluster collisions are calculated as special on-shell
matrix elements of transition operators $\mcu_{\beta\alpha}$ between
the channel states $| \phi_{\alpha}(\mbf{p}_\alpha) \rangle$
with the relative two-cluster momenta $\mbf{p}_\alpha$.
The primary interest of the present work are the atom-trimer and dimer-dimer scattering lengths
\begin{subequations} \label{eq:a}
\begin{align}  \label{eq:a13}
  A_{13} = {}& 3 \pi (3m/4)  \langle \phi_1(0)| \mcu_{11}|\phi_1(0) \rangle |_{\mcj=0}, \\
  A_{22} = {}& 2 \pi m  \langle \phi_2(0)| \mcu_{22}|\phi_2(0) \rangle |_{\mcj=0}, \label{eq:a22}
\end{align}
\end{subequations}
where the matrix elements are taken between states with vanishing relative two-cluster momentum and
total four-body angular momentum $\mcj=0$ at energies $E = -B_3$ (atom-trimer)
and $-2B_2$ (dimer-dimer).

 Momentum-space partial-wave basis 
 $|k_x k_y k_z [(l_x l_y)J l_z] \mathcal{JM} \rangle_\alpha$ is used to solve AGS equations (\ref{eq:U})
 where they build a system of coupled integral equations with three continuous variables
 $k_x, k_y, k_z$, the  magnitudes of  Jacobi momenta \cite{deltuva:07a}. The
 associated orbital angular momenta $l_x, l_y, l_z$ via $J$
 are coupled to $\mathcal{J}$ with the projection $\mathcal{M}$. 
 Discretization of three Jacobi momenta results in a large system of linear algebraic equations.
More details on the solution methods are given in  Ref.~\cite{deltuva:07a}.

\section{Results for the four-atom system \label{sec:res}}

\subsection{Tetramer ground state energy}

\begin{figure}[!]
\begin{center}
\includegraphics[scale=0.64]{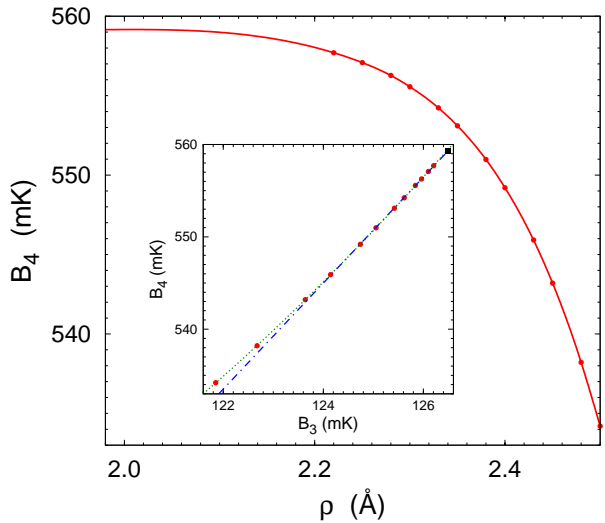}
\end{center}
\caption{\label{fig:b4} (Color online)
\He tetramer ground state binding energy $B_4$ as a function of the softening parameter $\rho$.
The curve is obtained by the extrapolation in the variable $B_3$.
The inset shows the $B_4$ dependence on the trimer ground state binding energy $B_3$.
The dashed-dotted line represents the linear fit to data points 
2.2 \AA{} $ \le \rho \le 2.43$ \AA{}, while the dotted curve 
represents the quadratic fit to all data points up to $\rho = 2.5$ \AA.
Extrapolation to the $B_3$ prediction of the original LM2M2 potential is given by
the black square.
}
\end{figure}

To validate the proposed SE method in the four-atom system I start with the calculation of the
tetramer ground state binding energy $B_4$ where several well established results are available
\cite{blume:00a,lazauskas:he,das:11a,hiyama:12a}. The Faddeev-Yakubovsky equations (\ref{eq:fy})
are solved including partial waves with orbital angular momenta
$l_x, l_y, l_z \le 8$ and about 70 to 100 grid points for the discretization of Jacobi momenta;
trimer binding energy is calculated using the same model space.
The results are obtained for 2.2 \AA{} $ \le \rho \le 2.5$ \AA{}; for brevity the dependence 
of binding energies on $\rho$ is suppressed in the notation.
Figure \ref{fig:b4} shows the dependence of the tetramer ground state binding energy on $\rho$, indicating
that the extrapolation  in $\rho$ is problematic.
Some particular function of  $\rho$ could be more suitable. As  the inset of Fig.~\ref{fig:b4}
demonstrates, extrapolation in $B_3$ works very well since $B_4(B_3)$ in the considered regime is nearly a linear
function, as should be expected in the context of Efimov physics \cite{naidon:rev,kievsky:21arnps}
and was shown for \He atoms in Ref.~\cite{hiyama:12b}.
However, a slight deviation from the linearity is seen for  $\rho \ge 2.45$ \AA.
A probable reason is that tetramer, being more compact and therefore more sensitive to the short-range force than
trimer, starts to "feel" the softening of the barrier at lower $\rho$ values.
Nevertheless, a high-quality fit is obtained by the inclusion of the quadratic term, as shown by the dotted curve.
Alternatively, a linear fit (dashed-dotted curve) in the reduced region 2.2 \AA{} $ \le \rho \le 2.43$ \AA{}
works also well. In both cases the extrapolation to the limit of the original LM2M2 potential 
yields tetramer ground state binding energy value $B_4 = 559.3(1)$ mK.
This result is in good agreement with the most accurate available prediction $B_4 = 559.22$ mK
obtained using the variational Gaussian expansion method \cite{hiyama:12a}.
The results obtained with  other methods \cite{lazauskas:he},
perturbatively corrected for the updated $\hbar^2/m$ value
as in Ref.~\cite{hiyama:12a}, range from 557.2 to 559.3 mK. The observed
agreement confirms the reliability of the proposed SE method.

\subsection{Dimer-dimer scattering length}

\begin{figure}[!]
\begin{center}
\includegraphics[scale=0.64]{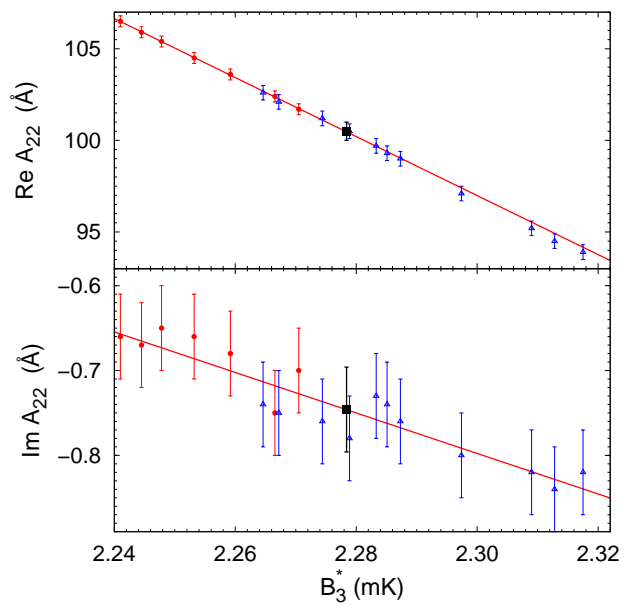}
\end{center}
\caption{\label{fig:a22} (Color online)
  Real and imaginary parts of the dimer-dimer scattering length  as functions of the binding energy $B_3^*$
  of the excited trimer state.
 Results obtained using the potential  (\ref{eq:vpr}) are shown as full circles, the lines are 
 linear fits to those results.
 Extrapolations to the $B_3^*$ prediction of the original LM2M2 potential are given by full squares.
 The points shown by  open triangles are obtained using  potentials of type (\ref{eq:vpr2})
and are not included in the fit.
}
\end{figure}

In contrast to $A_{12}$, the presence of a lower-lying threshold, atom plus trimer, renders
the dimer-dimer scattering length $A_{22}$ complex.
It is obtained by solving the AGS equations (\ref{eq:U12}) and (\ref{eq:U22}) at the energy $E=-2B_2$. 
Since the solution proceeds via the double Pad\'e method \cite{deltuva:07a},
its convergence and accuracy  are limited by a bad divergence of the Neumann series for four-body transition operators. This difficulty is
characteristic to Efimovian systems owing to a rich spectrum of states, corresponding to 
transition operator poles, that spoil down the convergence \cite{deltuva:11b}.
The results therefore are shown with their numerical error bars, typically below 0.5\% for the real part,
but above 5\% for the much smaller imaginary part.
Within those error bars
the  real and imaginary parts of  $A_{22}$ are consistent with the typical $\rho$-dependence,
already seen in the case of three-body
observables and tetramer binding energy; it is  therefore not presented here again. Instead,
Figure \ref{fig:a22} displays $\mathrm{Re}\,A_{22}$ and $\mathrm{Im}\,A_{22}$
as functions of the binding energy $B_3^*$
of the excited trimer state, the state closest to the two-dimer threshold.
On the other hand, given the linear  correlations demonstrated in Sec.~\ref{sec:3b},
one could chose $B_3$ or $A_{12}$ equally well without affecting any conclusions.
The calculated data points are shown as full circles, the lines are  linear fits to those results.
Extrapolation to the limit of the LM2M2 potential yields
\begin{equation} \label{eq:a22ri}
A_{22}  = [100.5(5) - i 0.75(5)] \, \mathrm{\AA}.
\end{equation}
Thus, the figure \ref{fig:a22} suggests the existence of the Phillips line also for the dimer-dimer scattering.
To strengthen this conclusion  additional calculations are performed using slightly different
scheme for the softening of the potential, namely
\begin{gather} \label{eq:vpr2}
  \langle p' | \bar{v}_L(\rho) | p  \rangle  =
  e^{-[\bar\lambda(\rho)p']^2} \, [1-\nu\bar\lambda(\rho)]\,
  ( \langle p' | V_L(\rho) | p  \rangle \, e^{-[\bar\lambda(\rho)p]^2}.
\end{gather}
$\bar\lambda(\rho)$ for each value of the parameter $\nu$ is again determined fitting the
dimer binding energy $B_2$. Introducing $\nu$ offers the flexibility to explore a broader
range of  $A_{22}$ and $B_3^*$, both above and below the LM2M2 value.
The additional data points in Fig.~\ref{fig:a22} represented by open triangles are obtained with
$\nu$ ranging from 0.1 to 0.5 \AA${}^{-1}$ and $\rho$ between 2.45 and 2.52 \AA.
Within the error bars the additional data points obtained using softened potentials of type (\ref{eq:vpr2})
are consistent with the linear correlation determined using the potential (\ref{eq:vpr}).

It is also interesting to compare the dimer-dimer scattering length (\ref{eq:a22ri}),  obtained for
the realistic potential,  with  the universal zero-range prediction.
From Fig.~2 of   Ref.~\cite{deltuva:11b} at the physical $B_3^*/B_2$ ratio one gets
$A_{22}/a \approx 0.945 - i0.004$. Thus, with the LM2M2 two-atom scattering length 
$a=100.0$ \AA{}  the zero-range limit $A_{22} \approx (94.5 -i0.4)$ \AA{} reproduces reasonably the real part
but fails significantly for the imaginary part. This is not very surprising since $\mathrm{Im}\,A_{22}$
is determined by the transition to the ground state trimer that is more affected by finite-range corrections.

The collision of two dimers  has an inelastic channel, i.e.,
rearrangement leading to the atom plus trimer state. 
In the gas consisting of dimers of the density $n_2$ this reaction would lead to the
trimer density $n_3$ increase in time as
\begin{gather} \label{eq:n3}
\frac{dn_3}{dt} = \beta_{22\to31} \frac{n_2^2}{2!}
\end{gather}
In the ultracold limit corresponding to
the vanishing relative dimer-dimer momentum $p_2 \to 0$
the inelastic cross section
$\sigma^0_{22\to31} \sim |\langle \phi_1(\mbf{p}_1)| \mcu_{12}|\phi_2(0) \rangle|^2 p_1/p_2  $
is formally infinite,
but the reaction rate  $\beta^0_{22\to31} ~\sim p_2 \sigma^0_{22\to31}$ is finite.
Using the  optical theorem it can be expressed via the imaginary part
of the dimer-dimer scattering length as
\begin{equation}  \label{eq:rlx0}
\beta^0_{22\to31} = -\frac{8\pi \hbar}{m} \, \mathrm{Im} (A_{22}).
\end{equation}
The numerical value is $\beta^0_{22\to31} = 3.0(2)\times 10^{-11} \, \mathrm{cm}^3/\mathrm{s}$;
the zero-range limit  would  underestimate it by a factor of two.

\subsection{Atom-trimer scattering and excited tetramer state}

Atom-trimer scattering calculations follow the same procedure as described in the
previous subsection for the dimer-dimer scattering except that AGS equations
(\ref{eq:U11}) and (\ref{eq:U21}) around the energy $E=-B_3$ are solved.
The extrapolation to the limit of the original LM2M2 potential yields the
\He atom-trimer scattering length $A_{13} = 108.8(5)$ \AA{} and the effective range
$R_{13} = 29.2(2)$ \AA. The latter agrees well with the only available prediction
$R_{13} = 29.1$ \AA{} in Ref.~\cite{lazauskas:he}. In contrast, there is nearly 5\% difference with
the prediction $A_{13} = 103.7$ \AA{} of Ref.~\cite{lazauskas:he} and a
strong disagreement with  $A_{13} = 56$ \AA{} of Ref.~\cite{blume:00a}.

The binding energy of the excited tetramer state $B_4^*$  is obtained 
looking for the energy corresponding to the pole of four-body transition operators
at  $E=-B_4^*$ in  Eqs.~(\ref{eq:U11}) and (\ref{eq:U21}). The extrapolation to the
LM2M2 point yields $B_4^* = 127.46(2)$ mK, or  $B_4^* - B_3 = 0.96(2)$ mK.
The weak binding of the excited tetramer state with respect to the ground state trimer
causes a number of methods to fail heavily on $B_4^* - B_3$,
despite quite accurate predictions for  $B_4$ \cite{blume:00a,das:11a}.
Even the two most advanced calculations differ by 15\%, with
$B_4^* - B_3 = 1.087$ mK in Ref.~\cite{lazauskas:he} and 0.93 mK in Ref.~\cite{hiyama:12a}.
The former is the estimation based on the effective-range expansion. The same approach using
$A_{13}$ and $R_{13}$ of the present work yields a smaller value $B_4^* - B_3 = 0.97(2)$ mK,
mainly due to a larger $A_{13}$ than in Ref.~\cite{lazauskas:he}.
Thus, the present results for  $B_4^* - B_3$ tend to support those of Ref.~\cite{hiyama:12a}.

\section{Summary and conclusions \label{sec:sum}}

Calculations of weakly-bound few-body atomic \He systems with realistic potentials are complicated due to
the very strong repulsion at short distances. While the three-body system is still manageable
using a number of methods, the four-body system requires a special treatment, especially in the
continuum. The present work proposed a softening and extrapolation approach for dealing with the short-range
repulsion and implemented it in the rigorous momentum-space framework for transition operators.

The strength of the short-range repulsion was gradually  reduced
by introducing one softening parameter,
at the same time suppressing the high-momentum components by a nonlocal extension of the potential,
adjusted to reproduce exactly the original dimer binding energy.
This ensured that also other few-body observables  deviate only mildly, within few percent,
from their original values. Furthermore, those deviations evolve smoothly with the softening  parameter,
allowing for the extrapolation back to the limit of the original potential.
Sufficiently accurate solutions of momentum-space three- and four-body
equations were obtained using the softened realistic LM2M2 potential.
Binding energies of trimer and tetramer ground and excited states
as well as the atom-dimer, atom-trimer, and dimer-dimer scattering lengths were calculated for a range of
softening parameter values. 
Nearly linear correlations between all those three- and four-body quantities were observed,
suggesting to use one of the easily calculable three-body quantities as the extrapolation variable,
thereby essentially improving the accuracy of the extrapolation procedure.
In particular, the linear correlation between the trimer binding energy and the dimer-dimer scattering
was demonstrated, establishing an additional case of the Phillips line.

The tetramer ground state energy, extrapolated to the limit of the original LM2M2 potential, agrees well
with previous determinations by other methods. In the more controversial case  of the excited tetramer state,
my prediction for its binding energy is less than 1 mK with respect to the atom-trimer threshold, clearly 
supporting the results of  Refs.~\cite{lazauskas:he,hiyama:12a}
over all the other \cite{blume:00a,das:11a}.
There is a reasonable agreement with Ref.~\cite{lazauskas:he} in the case of 
the atom-trimer scattering length and effective range, though my slightly larger $A_{13}$ value implies a
slightly weaker binding for the excited tetramer state, in a better agreement with Ref.~\cite{hiyama:12a}.

The most important result is the complex dimer-dimer scattering length. Its real part turns out to be very close
to the atom-atom scattering length, while the imaginary part is smaller by more than a factor of hundred.
The latter determines the rate of \He trimer production via two-dimer collisions in ultracold gases.
The universal zero-range theory is unable to reproduce accurately the imaginary part of the dimer-dimer
scattering length and the trimer production rate, thereby
indicating the importance of finite-range corrections.

\vspace{1mm}



\end{document}